\documentclass{article}

\usepackage{preprint}
\usepackage{upgreek}				
\usepackage[utf8]{inputenc} 		
\usepackage[T1]{fontenc}    		
\usepackage[dvipsnames]{xcolor}	
\usepackage{hyperref}       		
\hypersetup{
    colorlinks = true,
    linkcolor = {NavyBlue},		
    anchorcolor = .,
    citecolor = {NavyBlue},		
    filecolor = .,				
    menucolor = .,
    runcolor = {cyan}, 
    urlcolor = {NavyBlue},		
    }
\usepackage{url}            
\usepackage{booktabs}       
\usepackage{amsfonts}       
\usepackage{nicefrac}       
\usepackage{microtype}      
\usepackage{fancyhdr}       
\usepackage{graphicx}       
\usepackage{setspace} 
\usepackage{soul} 
\usepackage{ragged2e} 
\usepackage[toc,page]{appendix}
\usepackage{longtable}
\usepackage{eso-pic}
\usepackage{tikz}
\usepackage{siunitx}
\usepackage[superscript]{cite}

\graphicspath{{./Figures/}}     

  \centering
\title{Storing Sensor Events in the Interconnection Strength of Conducting Polymer Dendrites }

\author{
  Antoine Baron\textsuperscript{a}, Bilel Hafsi\textsuperscript{a,b},
  Enrique H. Balaguera\textsuperscript{c},
  S\'{e}bastien Pecqueur\textsuperscript{a}  \\
  \\
  a. IEMN, UMR 8520 \\
  Univ. Lille, CNRS, Univ. Polytechnique Hauts-de-France\\
  59000 Lille, France\\
  \\
  b. ICAM School of Engineering, \\
  6 rue Auber \\
  59800 Lille, France\\
  \\
  c. Escuela Superior de Ciencias Experimentales y Tecnolog\'{i}a, \\ Universidad Rey Juan Carlos, C/ Tulip\'{a}n, \\
  28933 M\'{o}stoles, Madrid, Spain\\
  \\
  \texttt{sebastien.pecqueur@iemn.fr} \\
}

\begin{document}
\maketitle

\begin{abstract}
If electronics drives only electrons to charge electrodes, natural systems learn by moving matter to evolve.
Morphogenesis in sessile organisms can be seen both as a fabrication process and as an operative mechanism.
However, intricating manufacturing and programming functionalities in electronic hardware is not conventional.
In this study, we experimentally implement such a concept of an evolutionary electrical system using a neuro-inspired electronic nose as a model, to store a history of sensory data in the physical properties of the electrical interconnects of sensing elements.
Triggered only by volatile molecule exposures, different sensing elements change instantaneously and reversibly their impedance, so pulse voltages enable the electrochemical growth of conducting polymer dendrites.
The strength of the evolving interconnects is specific to the sensing materials and to the nature of volatile molecules to which they are exposed.
The dendritic growths occur exclusively when exposed to volatile samples, and stop immediately after interrupting the exposure.
The capability of such "passive memory" was also assessed by simulating a network architecture, which showed that this way of storing information should greatly diminish the fabrication complexity of a highly dense sensing array while realistically enabling its calibration to classify user-specific environment exposures.
By demonstrating that memory in electronics can be a concept linked to manufacturing like in living organisms, this study shows that low material resources and low energy activation can be exploited for practical electronic applications in future-emerging sensing technologies.
\end{abstract}

\raggedright

\newpage 

\justifying

\section{Introduction}
Modern electronics consume increasing amounts of energy, especially due to data centers and connected devices, increasing the global carbon footprint.\cite{deVries2023,EPRI2024,IEA2025} 
Also, the massive collection of data by these devices raises serious privacy concerns, with higher risks of surveillance and personal data breaches.\cite{Hammouchi2019,Li2023,Gracy2025}
The core issue is largely tied to the materials themselves.\cite{Boyd2012}
Recycling silicon is highly complex and energy-demanding.\cite{Li2021,Preet2024,Duan2024,Murugesan2025}
Also, this element requires energy-intensive processes to be purified and crystallized as a starting material and to be structured as functional chips in foundries drawing large amounts of electricity.\cite{Wang2023,DiSabatino2024,Liu2024}
Its rigid and substrate-based nature allows only electrons to move, and because it is fabricated layer by layer, modern electronic systems remain fundamentally flat and still,\cite{Singh2024,Jayachandran2024,Padovani2024,Kim2025} unlike many naturally evolved intelligent systems, which are three-dimensional and dynamic.\cite{Stepanyants2002,Tero2010,Chaiwanon2016}
The inability of physical hardware to exploit the whole bulk of a silicon die, or to be structurally reconfigurable is a major limitation to process large amounts of data in real time.\cite{Yao1999,Zebulum2001,Haddow2011} 
As a result, IoT devices must rely either on oversized onboard computational resources or on remote data transmission to access more powerful server-side processing.\cite{Andriulo2024,Ali2024,Huang2025}\\[3pt]
To build truly efficient systems, we need to better understand the fundamental properties they require. 
Biology provides valuable insights, as living organisms achieve both fabrication and information processing through low-energy, integrated processes, without necessarily sharing data to perform intelligent tasks.
The human brain illustrates such efficiency: it consumes around 20~W while performing complex recognition tasks locally,\cite{Laughlin2003,Balasubramanian2021} relying on rich sensory inputs rather than constant communication.\cite{Purves2003} 
Overall, living systems are chemical machines. 
They sense and process information through molecular interactions, enabling capabilities such as olfaction, that remain difficult to replicate artificially.\cite{Bushdid2014,Zhang2026}
Even at the synaptic level, chemical signaling allows multiple messages to be encoded simultaneously and in a single junction,\cite{ORourke2012,Oostrum2025} whereas silicon-based systems are largely limited to flows of electrons only.\\[3pt]
Another key feature of biology is morphogenesis.\cite{Jan2010,Arikkath2012,Prigge2018,Lefebvre2021} 
Living organisms do not grow structures arbitrarily.
Growth is a way to process information patterns by itself.\cite{Collinet2021,Coen2023} 
For instance, neural dendrites develop based on experience,\cite{Xu2012,Voigts2019,d’Aquin2022} forming task-specific topologies adapted to particular functions. 
This contrasts sharply with silicon: rigid and static at ambient conditions.
Instead of relying on fixed, layer-by-layer fabricated substrates, we could explore alternative materials that enable bio-inspired computational systems capable of sensing chemical information from their environment to drive the growth of adaptive evolving circuits.\cite{Greenwood2006,Greenwood2015}\\[3pt]
To this aim, conducting polymers are promising materials to enable sensitive and transient electronics:\cite{Erokhin2007,Smerieri2008}
First, conducting polymers can be sensitive to different physical properties such as light,\cite{JansenvanVuuren2016,FuentesHernandez2020,Zhang2023} temperature,\cite{Bharti2018,Zhao2020,Xu2021} or pressure,\cite{Pan2014,Yang2019,Zong2025} but also chemicals like ions and neutral molecules.\cite{Janata2003,Lange2008,Liu2023,Dube2025}
Their ability to tune their chemical sensitivity in a versatile way benefits electrode arrays for machine-learning driven identification on electronic noses (e-Noses) and e-Tongues.\cite{Hatfield1994,Freund1995,Schaller2000,Barisci2002,Lewis2002,Gruber2013,Tiggemann2017,Park2019,SierraPadilla2021}
Second, conducting polymers can form dendrites (CPDs) whose morphologies promote signal classification.\cite{Cucchi2021,Cucchi2021a,Scholaert2025}
As they can serve both as interconnects and as transistors,\cite{Curtis1993,Sailor1994,Petrauskas2021,Janzakova2021,Scholaert2022} their complex topologies induce morphology-dependent programmability.\cite{Janzakova2023,Baron2025}

\begin{figure}[!h]
  \centering
  \includegraphics[width=1\columnwidth]{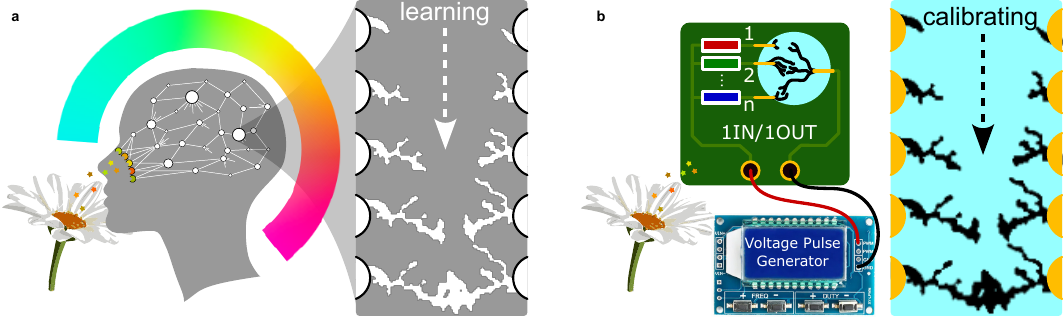}
  \caption{\textbf{Core Concept of this Study $\vert$}
  \textbf{a,} The brain using dendrite morphogenesis to learn classifying different olfactory perceptions from the input of a variety of receptor neurons.  
  \textbf{b,} In this study is implemented CPD morphogenesis to calibrate an electrical circuit composed of sensing elements, to respond to the exposure of volatile molecules by the growth of dendritic interconnections between a sensing input layer and an output, only by the energetic supply of low voltage pulses and electroactive molecules.
  }
  \label{fig:fig1}
\end{figure}

Third, CPDs grow and evolve strictly via electrochemistry:\cite{Koizumi2016,Ohira2017,Watanabe2018,Koizumi2018,Chen2023,Villani2023} With a minimal amount of chemical energy stored in a solution as a wet substrate,\cite{Baron2026} fractal structures expand in a 3D space over time, triggered with small voltages in ambient conditions.\cite{Janzakova2021}
Like their biological counterpart, CPDs conduct ions but interact with electrons in the material's bulk.\cite{Rivnay2016,Paulsen2020,Gkoupidenis2024}
Because different morphologies induce different charge time constants,\cite{Baron2024,Baron2024a} the structural evolution of CPDs induces adaptability to an electrical circuit during operation, shifting to a manufacturing paradigm where circuits transform during their use,\cite{Greenwood2010,Biswas2016,Cheng2016,Fu2016,Daehn2018} at low voltage and with restricted amount of chemical energy.\\[3pt]
This study demonstrates a first practical implementation of such an evolving circuit using an impedimetric eNose as a model,\cite{He2020,Lu2021,daSilva2025} supervised to be specifically sensitive to user-defined environments by mimicking the brain's topological plasticity (Fig.~\ref{fig:fig1}).\\[3pt]

\section{Experimental}

\subsection{Materials and methods}
All chemicals were used without further purification and were exposed to air, ambient light and room temperature. 
3,4-Ethylenedioxythiophene (EDOT), sodium polystyrene sulfonate (NaPSS, 70~kDa) and parabenzoquinone (BQ) were purchased from Sigma Aldrich. 
Two sensing materials, sodium poly[2-(3-thienyl)-ethoxy-4-butylsulfonate] (PTEBS -- from SolarisChem as SOL4140) and iron triflate doped PTEBS (Fe(OTf)\textsubscript{3}:PTEBS), which are sensitive to volatile molecules, were deposited via drop-casting onto pairs of concentric spiral electrodes (similar geometry as in previous studies on silicon)\cite{Pecqueur2018,Boujnah2021}, printed on FR-4 printed circuit boards (purchased from Aisler).
The sensitivity of PTEBS and doped PTEBS was reported in a previous study,\cite{Koudjou2026} and results from mixed ion-electron conduction of conjugated polyelectrolytes.\cite{Zeglio2015,Zeglio2017,Wan2022,Chae2024}
For both CPDs growth and formulation of sensitive materials, water is used as a solvent. 
\\[3pt]

\subsection{Electropolymerization Setup}
A bipolar square-wave signal of 80~Hz frequency and 50\% duty cycle was used for the electropolymerization of CPDs.
The voltage waveform was generated with a HW-753 PWM Signal Generator Module which allows voltage amplitudes up to 15~V\textsubscript{p}. 
The DC voltage was generated by a Keysight E36313A DC Power Supply.
The growth was performed on 25~$\upmu$m diameter gold wires purchased from Goodfellow, which were placed on micromanipulators to control the distance between the two wire electrodes, which in all experiments is set to 240~$\upmu$m. 
When three wires were used, they were oriented every 120° and equidistant from each other, so as to avoid introducing asymmetry in the impedance between inputs and output.
The growth was observed from above using a VGA CCD color camera from HITACHI Kokusai Electric Inc.

\subsection{Exposure Protocol}
Three beakers were filled at the same level with their respective volatile solvent (acetone, ethanol, and water). 
Five consecutive impedance measurements were performed after the breadboard holding the sensor was flipped at the top of a beaker, with the sensor directly facing the content of the beaker as pictured in Fig.~\ref{fig:fig4}d.
When two sensitive elements were used, both shared the same breadboard which was flipped above the water beaker and left 30~s at rest before applying the electropolymerization signal to prevent any transient response of a sensitive element from impacting the CPD growth.

\subsection{Electrical Characterization}
Characterization was performed \textit{in situ} without removing the dendrites from the solution droplet in which they were grown, eliminating the need for complex lithography steps.
The impedance measurements were performed using a Solartron Analytical (Ametek) impedance analyzer directly between the two wires without involving the rest of the circuit (sensors or ohmic resistor). 
A Rhode\&Schwarz RTB2004 digital oscilloscope was used to measure the voltage between the two gold wires. The raw voltage signals are first smoothed using a centered moving average with a 50-point window to filter out high-frequency noise, and subsequently decimated by a factor of ten. After shifting the time axis to synchronize all measurements, the upper and lower signal envelopes are extracted by identifying the absolute maximum and minimum voltage values within fixed time intervals of 12.5~ms.\\[3pt]
\newpage

\newpage

\section{Results}

\subsection{Effect of a Resistive Load on CPD Morphogenesis}
\begin{figure}[h]
  \centering
  \includegraphics[width=1\columnwidth]{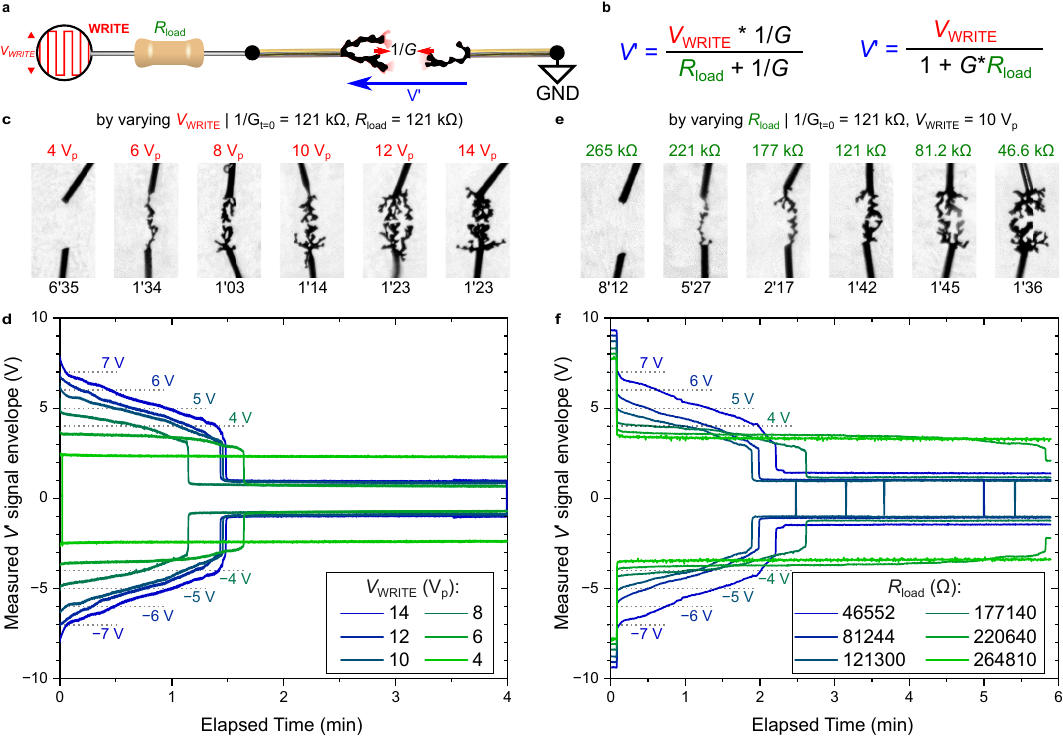}
  \caption{\textbf{Influence of the Growth Voltage Magnitude and Resistance Value of a Dummy Resistor on CPD Growth $\vert$ }
  \textbf{a,} Schematic of the circuit used in this part with its relevant voltages. 
  \textbf{b,} Expressions of the voltage V' based on the resistance value R\textsubscript{load} and the electrolyte resistance. 
  \textbf{c,} Highly contrasted optical microscopy pictures of CPDs grown with increasing \textit{V}\textsubscript{WRITE}, with an initial electrolyte solution resistance equal to 121~\si{\kilo\ohm} and a resistance value set to 121~\si{\kilo\ohm}. 
  \textbf{d,} Evolution of the envelopes of \textit{V}' over time during the growth for each CPD shown in \textbf{c}.
  \textbf{e,} Highly contrasted optical microscopy pictures of CPDs grown with increasing resistance values, an initial electrolyte solution resistance equal to 121~\si{\kilo\ohm} and \textit{V}\textsubscript{WRITE} set to 10~V\textsubscript{p}. 
  \textbf{f,} Evolution of the envelopes of \textit{V}' over time during the growth for each CPD shown in \textbf{e}.
  }
   \label{fig:fig2}
\end{figure}

The impedance of a serial sensitive element will determine whether dendritic growth is enabled by promoting a sufficiently high voltage \textit{V}' across the electrochemical cavity on both wires,\cite{Janzakova2021} as a voltage divider (Fig.~\ref{fig:fig2}a--b). 
Considering such a passive element to be voltage linear, a simple ohmic resistor of value \textit{R}\textsubscript{load} could serve as an "insensitive element" model (assuming no frequency shift).
The input voltage \textit{V}\textsubscript{WRITE}, a square-wave growth signal as defined in previous studies.\cite{Baron2025,Baron2026} is applied between the serial system composed of a resistive load and the electrochemical cavity. 
The resistance of the electrochemical cavity defined by the immersed wire system is given by the ion conductance \textit{G}. 
The voltage divider created by the series connection of the resistor and the electrolyte resistance 1/\textit{G} results in the expressions of \textit{V}' given in Fig.~\ref{fig:fig2}b. 
To verify such dependency for the CPD growth with \textit{V}\textsubscript{WRITE} and \textit{R}\textsubscript{load}, two series of experiments were carried out:\\[3pt]
First, the impact of \textit{V}\textsubscript{WRITE} was studied with a fixed value of \textit{R}\textsubscript{load} equal to 1/\textit{G} (Fig.~\ref{fig:fig2}c--d), dividing the input voltage by a factor of two (the \textit{R}\textsubscript{load} value is assumed from impedance spectroscopy). 
The peak voltage amplitude for \textit{V}\textsubscript{WRITE} was varied between 4~V\textsubscript{p} and 14~V\textsubscript{p}. 
The obtained CPD morphologies for each voltage are presented in Fig.~\ref{fig:fig2}c (pictures correspond to the CPDs state one second after completion, or at the end of the experiment).
As expected, the presence of a serial resistive load shifts up the voltage window to promote CPD growths.
Earlier results of Janzakova and coworkers in identical electrochemical conditions showed that a minimum of 3.5~V\textsubscript{p} was required to promote a growth, while 7~V\textsubscript{p} promoted bubbling at the gold wires.\cite{Janzakova2021}

In our experiments, we observed that 4~V\textsubscript{p} was not sufficient to trigger the growths, while no bubbling was observed at 14~V\textsubscript{p} (Fig.~\ref{fig:fig2}c).
No significant dependence of V\textsubscript{p} on the growth velocity was observed in the case of placing a resistive load (Fig.~\ref{fig:fig2}c), unlike what was observed by Janzakova and coworkers without a resistive load.\cite{Janzakova2021}
Still, we observed that the growth morphology was highly dependent on the voltage amplitude: higher voltages promoting higher dendricity for the CPDs (Fig.~\ref{fig:fig2}c).
When measuring the envelope of the divided voltage \textit{V}', we confirm the monotonic dependence of the envelope with \textit{V}\textsubscript{WRITE}, despite no strict reduction by a factor of two was observed (Fig.~\ref{fig:fig2}d).
For 4~V\textsubscript{p}, the envelope signal is steadily at 3.5~V in absolute value (Fig.~\ref{fig:fig2}d).
For higher V\textsubscript{p}, the envelope starts at an absolute value which is sufficient to promote a CPD growth.
The decrease over time is characteristic of the 1/\textit{G} diminishing, caused by the growth of material which increases the characteristic surface of the electrodes and reduces the characteristic gap between them.
At the time CPDs contact on the camera, a sudden reduction of the envelope at about 1~V in absolute value is observed (Fig.~\ref{fig:fig2}d).
The fact that this value appears to be poorly dependent with V\textsubscript{p} indicates that the CPD morphology has only moderate impact on the signal transport, once the CPD has reached its most mature state.\\[3pt]
Second, a similar study was conducted by fixing the value of \textit{V}\textsubscript{WRITE}'s peak voltage to 10~V\textsubscript{p} (sufficiently high to promote a growth without bubbling when \textit{R}\textsubscript{load} equals 1/\textit{G} at start) and varying \textit{R}\textsubscript{load} from 46.6~\si{\kilo\ohm} (about 0.4/\textit{G} at start) to 265~\si{\kilo\ohm} (about 2.2/\textit{G} at start --- Fig.~\ref{fig:fig2}e--f), dividing the input voltage by different values.
An opposite trend of the evolution of the CPD morphology is observed with the variation of \textit{R}\textsubscript{load} as with \textit{V}\textsubscript{WRITE} previously (Fig.~\ref{fig:fig2}e).
By its effect on the divided voltage \textit{V}' applied across the electrochemical cavity, a large value for \textit{R}\textsubscript{load} tends to disable the growth, while its decrease promotes higher dendricity (Fig.~\ref{fig:fig2}e).
A dependency on the completion voltage with \textit{R}\textsubscript{load} was observed: smaller resistive loads tend to increase the growth velocity (Fig.~\ref{fig:fig2}e).
The envelope of the divided voltage \textit{V}' displays also an opposite trend with \textit{R}\textsubscript{load} compared to \textit{V}\textsubscript{WRITE} (Fig.~\ref{fig:fig2}f).
These results confirm quantitatively the influence of a serial resistive load on the CPD growth: that system's supplied voltages should be increased to compensate for the generated voltage drop to observe the voltage-dependent morphogenesis, and that for a fixed voltage supply, any variation of the serial load should directly impact a growth morphology.\\[3pt]

\subsection{Competition of Parallel Loaded CPDs}
\begin{figure}[h]
  \centering
  \includegraphics[width=1\columnwidth]{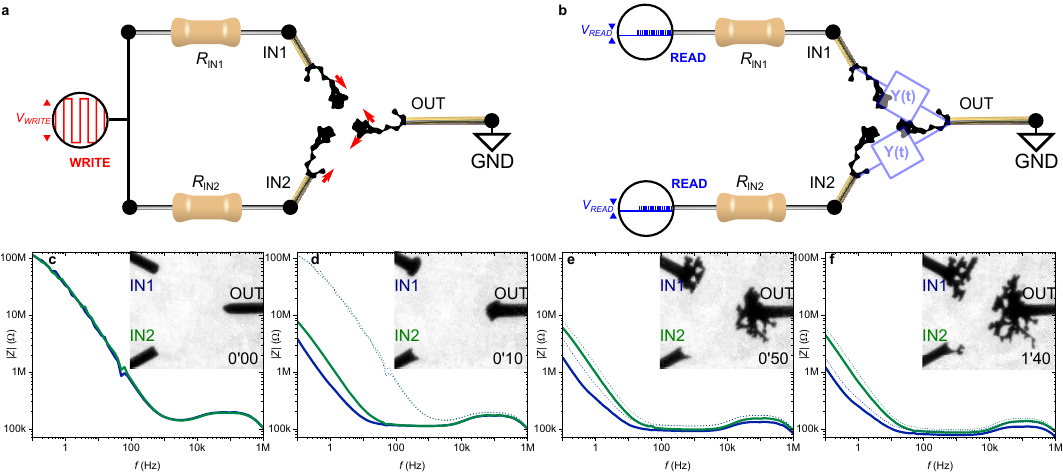}
  \caption{\textbf{Simultaneous Growth of Two CPDs from Two Different Sensors Emulated by Ohmic Resistors $\vert$ }
  \textbf{a,} Schematic of the circuit used in this part, in the writing phase where a high-voltage alternating signal is simultaneously applied at both input wires. 
  \textbf{b,} Schematic of the same circuit in the reading phase, where low-voltage pulsed signals can be sent separately to each input wire. 
  Each CPD can then be read as their admittance contains information about the impedance in series with them at the time of their growth. 
  \textbf{c--f,} Impedance measurements along the two paths at four stages of the CPD growth: before growth (\textbf{c}), after 10~s of growth (\textbf{d}), after 10+40~s of growth (\textbf{e}), and after 10+40+50~s of growth (\textbf{f}).
  Corresponding microscope pictures of the system are included as insets.
  }
   \label{fig:fig3}
\end{figure}

Dendrites in a common electrolyte can influence each other's conductances, behaving as mutually-gated transistors.\cite{Cucchi2021,Janzakova2021b,Scholaert2022,Scholaert2025} 
Their growth behavior when growing in parallel in a resistance-loaded circuit can therefore differ from the behavior of single growing CPDs.
To assess the competing effect of multiple growths arising from many sensitive elements in a common electroactive solution, CPD growths were studied with parallel systems of serial resistive loads (Fig.~\ref{fig:fig3}).
A circuit involving two ohmic resistors connected at the input wires to a single output wire is depicted in Fig.~\ref{fig:fig3}a--b. 
The same writing signal is applied to both resistors to trigger a simultaneous growth of two CPDs along the input pathways (Fig.~\ref{fig:fig3}a). 
Instead of reading both \textit{V}' envelope signals \textit{in operando}, the interconnectivity assessment was made intermittently after several write-and-read cycles, as in previous studies (Fig.~\ref{fig:fig3}b).\cite{Baron2024,Baron2025} 
The readout must be performed with a low voltage amplitude, so reading the system does not modify the junction state by promoting further electrodeposition or damaging the polymer.
Here, this readout was performed via impedance spectroscopy,\cite{Baron2024} to evaluate the electrical properties of each electrochemical junction.
The experiment was conducted with three growth interruptions to compare the interconnection asymmetry at four growth stages.
The asymmetry was promoted with two different resistive load values: 46.6~\si{\kilo\ohm} on IN1 and 177~\si{\kilo\ohm} on IN2.
These values were chosen low enough to promote growth simultaneously on all three wires, but different enough to observe noticeable morphology differences.\\[3pt]
In our experiments, we observed that the CPD growth on OUT is systematically bulkier at any maturity stage than each growth on the IN side (Fig.~\ref{fig:fig3}c--f).
As Janzakova and coworkers showed, this is not absolute and it can be conveniently tuned by adding a DC voltage component in the voltage signal: either with a voltage offset or with a different duty cycle in a pulse waveform.\cite{Janzakova2021}
In this specific case with no DC component in the voltage signal, the OUT CPD mass should equal the sum of the ones on all INs.
As the resistive load is lower on IN1 than on IN2, the CPD growing on IN1 is more massive than the one on IN2 at any stage (Fig.~\ref{fig:fig3}c--f).
Despite the load asymmetry, one IN does not disable the other one: the input side can be thought as a single electrode with different potential landscape where growths occur in parallel without interaction of one input on the other (the ohmic loads do not delay the signal).

As a consequence, input CPDs grow towards the one on the output and no directional growth between IN1 and IN2 is observed (Fig.~\ref{fig:fig3}c--f).
With the morphology asymmetry, significant impedance differences are recorded during the whole evolution of the system. 
The first graph shows that both paths have roughly the same impedance initially, as PEDOT:PSS is not coated on the naked gold electrodes (Fig.~\ref{fig:fig3}c). 
We noticed an impedance modulus increase from 2~kHz which has not been previously observed on a two-wire setup.\cite{Baron2024,Baron2024a}
After 10~s of growth (Fig.~\ref{fig:fig3}d), IN1 connected with the less resistive resistor by growing a more developed CPD, which features a lower impedance than IN2 at low frequency, characteristic of the CPD growth.\cite{Baron2024a,Baron2024}
In an electrochemical system composed of a pair of CPDs, the system's impedance is theoretically limited by the CPD which is the least bulky.\cite{Baron2024}
Because any CPD growing on an input is systematically smaller than the one on the output, any impedance between an input and an output should therefore be limited by the morphology growing on the IN electrode.
After further growth with the amplification of morphological differences (Fig.~\ref{fig:fig3}e--f), both impedances continue to decrease as their respective IN CPD grows. 
As dendrites evolve in parallel, both towards less juvenile states, their impedance difference seems to slightly increase over time.
At any time in their development, a resistor mapping can be read as an impedance mapping of a dendritic culture, where each resistance value is stored in the growth state when fed by a common voltage supply.
Overall, this experiment shows that the state of an array of resistors can be projected as a morphology of growing CPDs in a common electroactive solution with a single voltage source.

\subsection{From Resistive Loads to Vapor Sensitive Elements}

\begin{figure}[h]
  \centering
  \includegraphics[width=1\columnwidth]{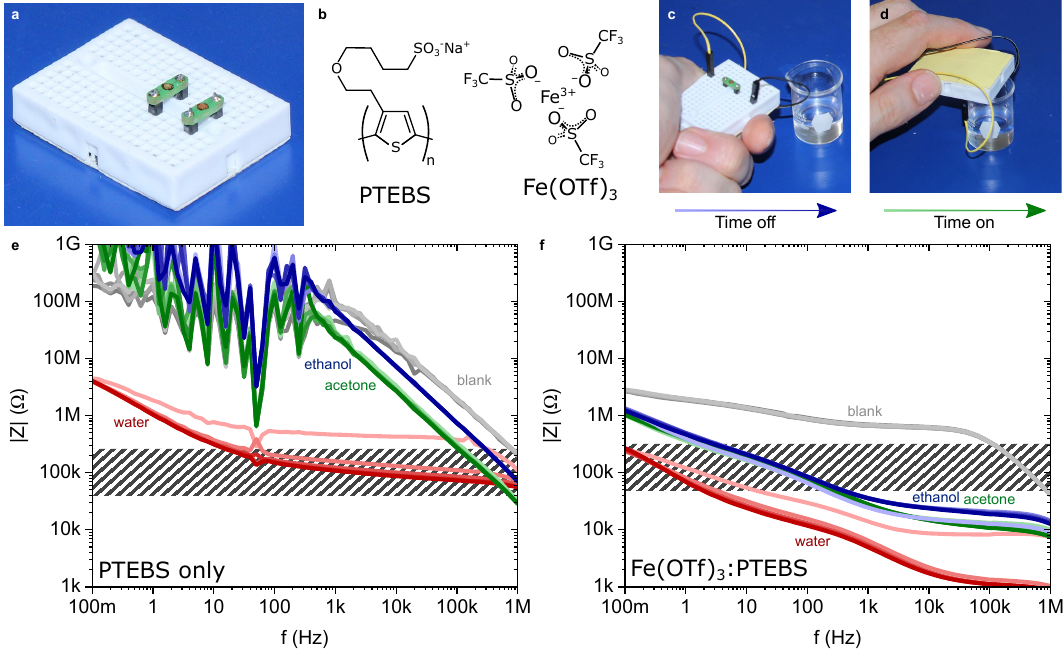}
  \caption{\textbf{Impedance Characterization of PTEBS Sensitive Elements $\vert$ }
  \textbf{a,} Picture of the two studied PTEBS elements on a breadboard. 
  \textbf{b,} Chemical formulae of PTEBS (left) and Fe(OTf)\textsubscript{3} (right). 
  \textbf{c,} Picture of the unexposed system in the "blank" state (OFF). 
  \textbf{d,} Picture of the exposed system to either water, acetone or ethanol states (ON). 
  \textbf{e,} Impedance spectra of the pristine PTEBS when exposed to the three volatile species and in the OFF state. 
  Each measurement contains five consecutive spectra to reach the steady regime. 
  The black-hatched area corresponds to the resistance values tested with the ohmic resistors in the previous parts. 
  \textbf{f,} Impedance spectra of the PTEBS doped with Fe(OTf)\textsubscript{3} when exposed to the three volatile species and in the OFF state.
  }
   \label{fig:fig4}
\end{figure}

As the value of a resistive load impacts the morphology of a growing CPD (Fig.~\ref{fig:fig2}) and that specific imprinted morphologies engrave specific impedance profiles on the electrochemical cavity (Fig.~\ref{fig:fig3}), we hypothesize that replacing such a load by a time-variant sensitive element should allow imprinting an environmental activity on the morphology of such growth.
In the next experiments, we consider such sensitive elements made of conducting polymers (specifically conjugated polyelectrolytes), either doped or not with a triflate salt (Fig.~\ref{fig:fig4}).\\[3pt]
This type of material has been recently studied by Koudjou \textit{et al.} for their potential to widen the receptive field of an eNose by diversifying the nature of doping salts.\cite{Koudjou2026}
Materials are simply drop casted on individual pairs of electrodes, to constitute individual sensitive elements which can be assembled into a system on a breadboard (Fig.~\ref{fig:fig4}a).

In this experiment, we limited our study to two materials: PTEBS as a conducting polyelectrolyte, which is either doped or not with Fe(OTf)\textsubscript{3} (chemical structures in Fig.~\ref{fig:fig4}b).
The sensitive elements on a breadboard are not integrated with the electrochemical cavity, and both are simply connected externally with cables.
The fact that both parts of the system are physically separated allows exposing the sensitive elements to different volatile chemical environments (Fig.~\ref{fig:fig4}c--d), while preserving the CPD growth from the exposure of any volatile molecules potentially interfering with their growth.\cite{Baron2026a}
As shown by Koudjou and coworkers, doped conjugated polyelectrolytes are stable in air and their impedance changes reversibly with the exposure depending on the partial pressure of acetone, ethanol and water.\cite{Koudjou2026}
The impedance changes for the sensitive elements used in this study are displayed as Fig.~\ref{fig:fig4}e for the pristine PTEBS and Fig.~\ref{fig:fig4}f for PTEBS doped with Fe(OTf)\textsubscript{3}.
The different responses of these materials have been explained as an intricated mechanism of ion-sensitive mobility in the hydrophilic polymer matrix, where different electrophilicities provided by different metals induce extended sensitivity to organic molecules.\cite{Koudjou2026} 
As a result, the PTEBS element has only low impedance when exposed to water vapors (Fig.~\ref{fig:fig4}e).
The impedance decreases readily and reversibly, by orders of magnitude in impedance modulus |\textit{Z}|, so its impedance value matches the range of resistances previously studied to trigger the growth of CPDs.
In case of ethanol, acetone or nothing is exposed to the PTEBS element, the impedance is high, so none of these exposures are expected to trigger a growth.
The impedance profile appears widely different when Fe(OTf)\textsubscript{3} is coated on PTEBS (Fig.~\ref{fig:fig4}f).
First, the impedance is far more reduced compared to pristine PTEBS thanks to Fe(OTf)\textsubscript{3} as both a p-dopant for PTEBS and as an ion conductor in PTEBS (the mixed conduction is evidenced by a non-RLC behavior of the impedance profiles characteristic of diffuse conductions).
The material shows sensitivity to acetone and ethanol, such that both appear to lower the impedance deep-enough to trigger a CPD growth (water is still expected to enable it as well by the even lower impedance induced on the Fe(OTf)\textsubscript{3}:PTEBS element).\\[3pt]

\newpage

\subsection{Triggering a CPD Morphogenesis with Volatile Molecules}

\begin{figure}[h]
  \centering
  \includegraphics[width=1\columnwidth]{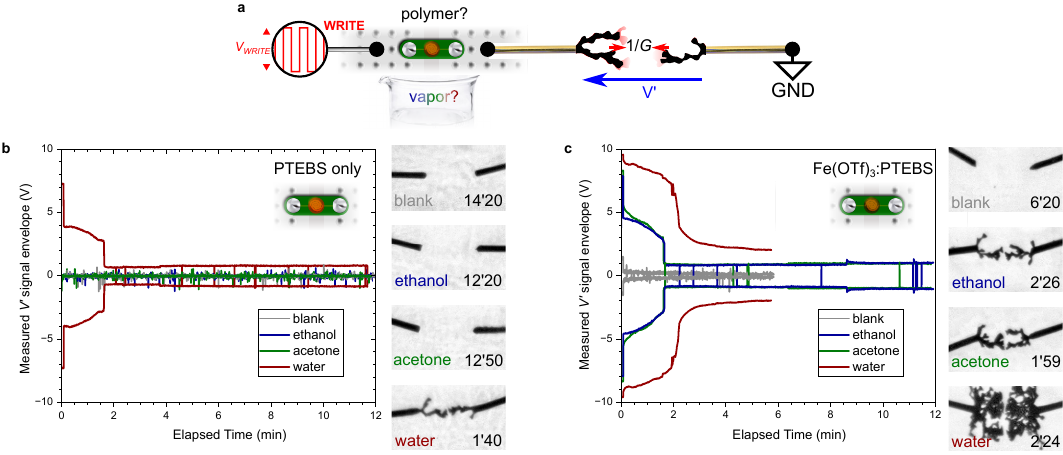}
  \caption{\textbf{Integration of CPDs with Sensitive Elements $\vert$ }
  \textbf{a,} Schematic of the circuit used in this part where the sensitive element is coated with a vapor sensitive polymer. 
  \textbf{b,} Evolution of the \textit{V}' signal envelope during CPD growth when pristine PTEBS is exposed to the different vapors (or in the OFF state). 
  \textbf{c,} Evolution of the \textit{V}' signal envelope during CPD growth with Fe(OTf)\textsubscript{3}-doped PTEBS.
  }
   \label{fig:fig5}
\end{figure}

In the following experiments, the integration of the CPDs with vapor sensitive elements was tested, using the electric circuit presented in Fig.~\ref{fig:fig2} but substituting the ohmic resistor with either the pristine PTEBS or the Fe(OTf)\textsubscript{3}:PTEBS sensitive elements that were characterized in Fig.~\ref{fig:fig4}.
A schematic is represented in Fig.~\ref{fig:fig5}a. 
The growth signal parameters remained identical to the one used in Fig.~\ref{fig:fig2} and the same exposure protocol used in Fig.~\ref{fig:fig4} has been replicated.\\[3pt] 

The exposures of the voltage divider comprising the pristine PTEBS element (results presented in Fig.~\ref{fig:fig5}b) confirm that no dendrite grows over a dozen of minutes of voltage polarization, unless the sensitive element is exposed over water.
Upon no exposure, or the exposure of acetone or ethanol, no changes (neither growth nor bubbling) were detected on the gold wires (Fig.~\ref{fig:fig5}b).
The \textit{V}' signal envelope in these cases records almost no signal applied between both wires during the 12~min experiment (Fig.~\ref{fig:fig5}b).
This results from the high impedance (previously shown in Fig.~\ref{fig:fig4}), causing the \textit{V}\textsubscript{WRITE} voltage to drop mainly through the sensitive element.
In the case of exposing the pristine PTEBS element to water vapors, a growth was observed (Fig.~\ref{fig:fig5}b).
Chronologically, the exposure of water was performed after the "blank", which indicates that the sensitive element was not damaged in air, under the stress with a quasi 10~V\textsubscript{p} voltage drop across it during 12~min.
In less than 2~min, a "177~\si{\kilo\ohm}" like CPD grows.
We notice that the decrease of the \textit{V}' signal envelope is less linear as previously observed with ohmic resistors in Fig.~\ref{fig:fig2}, which is attributed to the low transience of the element's impedance state when exposed to vapors.\\[3pt]
In the case of exposing the Fe(OTf)\textsubscript{3}-doped PTEBS element, a growth was observed in all cases except "blank" (Fig.~\ref{fig:fig5}c).
CPDs grown under acetone and ethanol vapor exposures of the sensitive element were completed in about two minutes with very similar "121~\si{\kilo\ohm}" like morphologies (Fig.~\ref{fig:fig5}c).
The CPD grown under a water vapor exposure of the sensitive element has a very ramified structure (more than the extreme "46.6~\si{\kilo\ohm}" case in Fig.~\ref{fig:fig2}), characteristic of the lower impedance (no bubbling was observed).
\textit{V}' signal envelopes are also very distinct between "water" and the two other exposures to acetone and ethanol: the voltage drops substantially more across the electrochemical cavity than the sensitive element in case of water (Fig.~\ref{fig:fig5}c).
Surprisingly, the CPD does not complete earlier for water than for ethanol and acetone.\\[3pt]
These results confirm two aspects:
First, a sensitive element can project its impedance state to the morphology of a growing CPD and the fact that the element has a non-negligible reactance in its impedance is not a limiting factor.
Second, that the amplitude of the modification of growth scales with the depth of the impedance diminishing.

\newpage

\subsection{Sensitivity Competition in the Morphogenesis of Parallel CPDs}

\begin{figure}[h]
  \centering
  \includegraphics[width=1\columnwidth]{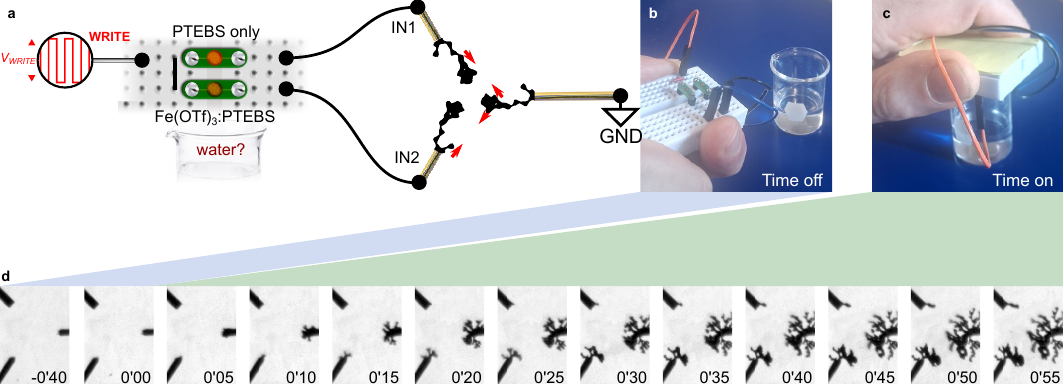}
  \caption{\textbf{Simultaneous Growths of CPDs with a Sensing Array Exposed to Water $\vert$ }
  \textbf{a,} Schematic illustrating the Y-shaped circuit used in this experiment.
  \textbf{b,} Picture of the system before exposing the sensors. 
  \textbf{c,} Picture of the system during the growth. 
  \textbf{d,} Optical microscope pictures showing the evolution of the CPDs during exposure to water.
  }
   \label{fig:fig6}
\end{figure}

In this last experiment, multiple sensitive elements are used as loads for two input wires immersed in the same electrochemical system.
A schematic is represented in Fig.~\ref{fig:fig6}a. 
Instead of testing the CPD growth triggered by many volatile molecules on a single sensitive element, the study exploits the sensitivity differences of pristine PTEBS and Fe(OTf)\textsubscript{3}:PTEBS to water vapors to verify if materials sensing a same chemical target can project different information on different growths, characteristic of their subtle specificity.
The study uses the electric circuit presented in Fig.~\ref{fig:fig3}, substituting the ohmic resistor with both the pristine PTEBS and the Fe(OTf)\textsubscript{3}:PTEBS sensitive elements that were characterized in Fig.~\ref{fig:fig4}.
The growth signal parameters remained identical to the one used in Fig.~\ref{fig:fig2} and the same exposure protocol used in Fig.~\ref{fig:fig4} has been replicated.

As both sensitive elements are exposed simultaneously to the water beaker (Fig.~\ref{fig:fig6}b--c), both input wires start their growth at the same time but with different morphologies (Fig.~\ref{fig:fig6}d).
The video recorded 40~s before this exposure and testifies that when a system is powered, no growth can occur if the impedance of the sensitive elements is not lowered by the presence of volatile molecules.
We noticed that air humidity did not influence the test.
Once the sensitive elements are exposed, a similar behavior as with the ohmic resistor occurred (Fig.~\ref{fig:fig3}).
Growth towards inputs and the single output is promoted without noticeable orientation of both inputs toward each other (Fig.~\ref{fig:fig6}c).
Also, the CPD on the single output wire is still more massive than the one on any of both input wires.
And, the CPD growth on the input wire connected to the Fe(OTf)\textsubscript{3}:PTEBS coated element is more ramified than the one on the input wire connected to the pristine PTEBS coated element.
Unlike the earlier experiment performed with the ohmic resistors displayed in Fig.~\ref{fig:fig3}, no impedance characterization was performed to not interrupt the growth intermittently in Fig.~\ref{fig:fig6}c at the risk of perturbing it with several intermittent vapor exposures on the sensitive elements.\\[3pt]
This experiment confirms that by modulating the impedance profile of different input nodes gathering materials with different sensitivities, the information projected on the CPD structures is also characteristic of the nature of the array, and not intrinsic to the environment.

\newpage

\subsection{Network Simulation to Calibrate 2\textsuperscript{n} Sensing Elements}
\begin{figure}[h]
  \centering
  \includegraphics[width=1\columnwidth]{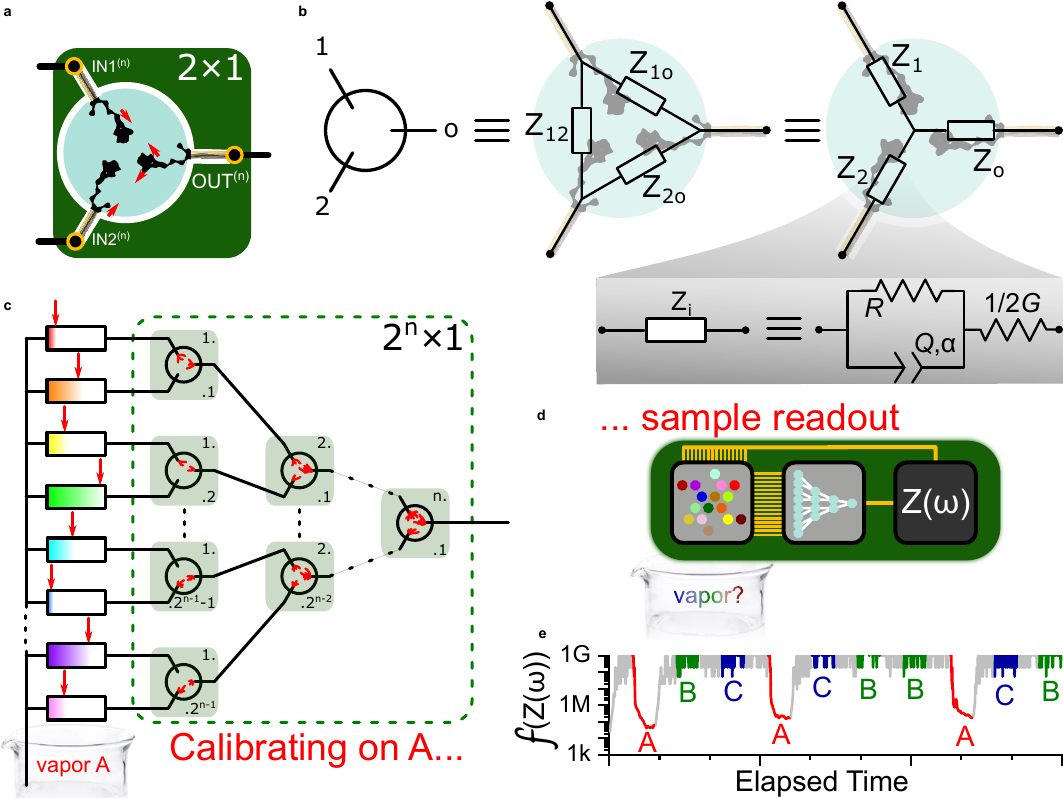}
  \caption{\textbf{A 2\textsuperscript{n}$\times$1 Block as a Network of Elementary 2$\times$1 Blocks to Calibrate a 2\textsuperscript{n}-Sensing Array $\vert$ }
  \textbf{a,} Proposing the Y-shaped circuit (characterized in Figs.~\ref{fig:fig3} and \ref{fig:fig6}) as an elementary block to connect two inputs to one output.
  \textbf{b,} Block symbol and proposed equivalent circuit for its operation below Faradaic regime (at readout).
  \textbf{c,} Generic expression of a 2\textsuperscript{n}$\times$1 as a binary tree structure of 2$\times$1 block elements to project the responses of 2\textsuperscript{n} sensing elements to one output terminal.
  \textbf{d,} Schematic of a proposed electrical hardware featuring exclusively a 2\textsuperscript{n} sensing array (passive), a 2\textsuperscript{n}$\times$1 block (passive) and a single input analyzer (active) as a bio-inspired electronic nose implementation.
  \textbf{e,} Expected output from the electronic nose, after being calibrated by A samples and no B and C samples (samples are not necessarily pure chemicals).
  }
  \label{fig:fig7}
\end{figure}

The former results successfully implemented artificial morphogenesis with conducting polymer electrochemistry to calibrate the impedance readout of an array composed of two sensitive elements.
The study was restricted to two inputs because equidistance between electrodes must be respected to not bias the growths towards closest neighbors, as the mechanism is field-activated.\cite{Koizumi2016,Ohira2017,Inagi2019}
Limited to two inputs with planar electrodes, conducting polymer filamentary switching with three inputs was achieved with three dimensional growths in a volume of solution.\cite{Hagiwara2023}
Although line passivation can increase the number of close neighbors to six inputs in 2D,\cite{Scholaert2025} and theoretically up to twelve inputs in 3D if arranged in a compact orthogonal crystal lattice,\cite{Pecqueur2025} these numbers are far too low compared to the number of receptor neurons a biological brain handles.
For example, human olfaction involves 3$\times$10\textsuperscript{6} olfactory receptor neurons,\cite{Moran1982} featuring 339 functional and specific receptors,\cite{Malnic2004} to classify complex molecular environments. 
At a deeper level of olfactory processing, neurons in the olfactory bulb project to distributed networks of neurons within the piriform cortex,\cite{Purves2003,Kandel2013} allowing us to project different odor perceptions.\cite{Kumar2015,Zhou2018,Lee2023} 
As a core mechanism of neural plasticity, dendrite morphogenesis is an essential mechanism responsible for the imprinting of new smell classes in our brain.\cite{Chou2010,Lyu2026}\\[3pt]
To apply CPD morphogenesis to address as many sensitive elements as a number of receptor neurons, the following strategy describes the use of the previously studied 2$\times$1 circuit in a network of them to address 2\textsuperscript{n} sensitive elements within a binary tree structure of 2$\times$1 subcircuits (Fig.~\ref{fig:fig7}a--b) to form a 2\textsuperscript{n}$\times$1 supercircuit (Fig.~\ref{fig:fig7}c).

\begin{figure}[h]
  \centering
  \includegraphics[width=1\columnwidth]{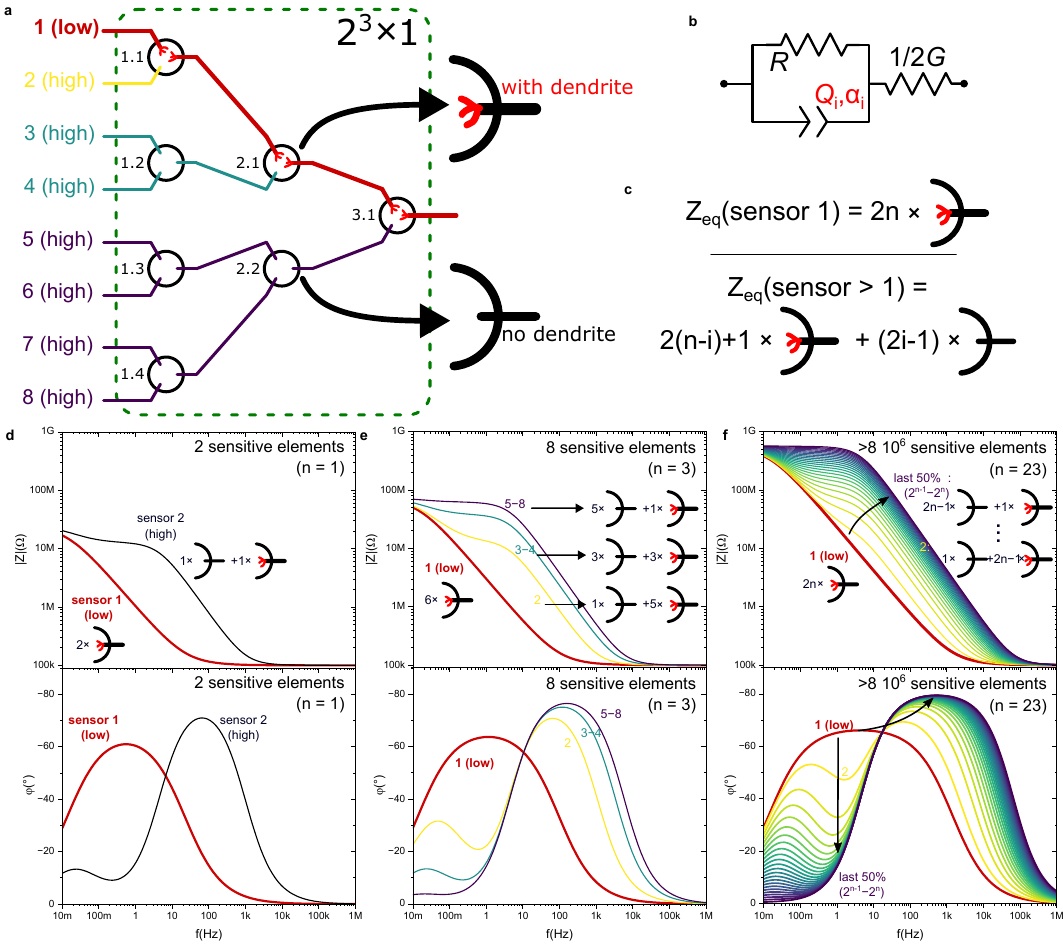}
  \caption{\textbf{Simulating the Impact of the Activation of One Single Element among 2\textsuperscript{n} in a 2\textsuperscript{n}$\times$1 Block after Calibration $\vert$ }
  \textbf{a,} A 2\textsuperscript{n}$\times$1 block (for n = 3) where only the sensitive element \#1 opens at calibration, creating a single path of grown CPDs.
  \textbf{b,} Equivalent circuit considered for simulating one wire/solution interface, where the discriminating parameters for an interface with and without a CPD are \textit{Q}\textsubscript{i} and $\alpha$\textsubscript{i}.
  \textbf{c,} Equivalent impedance for a path of a single input to the common output, defined by the number of grown CPDs.
  \textbf{d--f,} Simulated impedance modulus and phase for the different input-output paths, discriminating the impedance of the single "low" input from the many "high" outputs, for n = 1 (\textbf{d}), n = 3 (\textbf{e}) and n = 23 (\textbf{f}).
  }
  \label{fig:fig8}
\end{figure}

In an elementary 2$\times$1 subcircuit, the interconnectivity between two inputs and one output is strengthened for spike voltage activities that are the most intense in magnitude.
A 2$\times$1 three terminal junction is therefore a programmable comparator, with write-once/read-only memory which stores the past voltage difference between both inputs.
As a passive and analog circuit, the state of the memory can be accessed in the linear regime by applying small voltage amplitude either via impedance spectroscopy,\cite{Baron2024} or via transient current readout.\cite{Baron2025} 
The equivalent circuit of such electrochemical impedance is rather complex and is strongly morphology dependent.\cite{Baron2024,Baron2024a}
The simplest approximation would consist of a delta-circuit of equivalent impedances connecting the three couples of electrodes, equivalent to a Y-circuit where each element is a Randles subcircuit (without diffusion and considering a constant-phase element instead of an ideal capacitor) with a relaxation characteristic of the appended CPD (Fig.~\ref{fig:fig7}b).
At the level of a 2\textsuperscript{n}$\times$1 network, the circuit would have a more complex impedance as many CPDs grow at different levels in the network, each contributing in generating different harmonics (Fig.~\ref{fig:fig7}c).
When exposed to a specific environment, the tree structure of growing CPDs should specifically emerge from connections to the sensitive elements that exhibit comparatively the highest activity (Fig.~\ref{fig:fig7}c).
Thanks to such a classifier, a sensing hardware would not need a multiplexer stage to connect a sensing array to a readout circuit (Fig.~\ref{fig:fig7}d).\cite{Routier2024,Routier2024a}
Also, it would substitute traditional software classifiers such as Principal Component Analysis (PCA), which are used to prune an array and reduce the input dimensionality by silencing the data coming from the less relevant sensitive elements.\cite{Boujnah2022,HajAmmar2023,HajAmmar2024}
After calibrating in a specific environment, the system response is expected to imprint the characteristic impedance fingerprint, selectively connecting the sensitive elements which were opened during calibration to the output (Fig.~\ref{fig:fig7}e).\\[3pt]
To simulate the imprinted selectivity that such a hardware could potentially have, the multiple input-output impedances of different 2\textsuperscript{n}$\times$1 networks have been compared, in order to anticipate if the strategy is not limited at a specific number of layers n.
To do this, we simplified the approach to the case where calibrating to a single environment induces only one sensitive element in its "low" state (Fig.~\ref{fig:fig8}a), resulting in a single path of grown CPD from one input ("sensor 1") to the output.
As depicted in the figure \ref{fig:fig8}a, only the activated input is composed of serial elements "with dendrite", while the other non-activated elements gather a number of serial "without dendrite" elements, depending on the proximity of a given non-activated element has with the activated element in the tree structure.
The simplified simulation considers only a single state of completion for the CPD, so the impedance of an interface without CPD is associated with a subcircuit composed of a constant-phase element with parameters \textit{Q}\textsubscript{0} and $\alpha$\textsubscript{0}, while an interface without CPD is associated with \textit{Q}\textsubscript{1} and $\alpha$\textsubscript{1} (Fig.~\ref{fig:fig8}b).\\
As a serial circuit, the impedance Z\textsubscript{low} of the "sensor 1" in a 2\textsuperscript{n}$\times$1 network is (Fig.~\ref{fig:fig8}c):\\
\begin{equation}
Z_{low}(\upomega)=\frac{2nR}{1+(jRQ_1\upomega)^{\upalpha_1}}+n/G
\end{equation}
and the different impedances Z\textsubscript{high} of the other "sensor" in a 2\textsuperscript{n}$\times$1 network is (Fig.~\ref{fig:fig8}c):\\
\begin{equation}
Z_{high,i}(\upomega)=\frac{(2(n-i)+1)R}{1+(jRQ_1\upomega)^{\upalpha_1}}+n/G+\frac{(2i-1)R}{1+(jRQ_0\upomega)^{\upalpha_0}}
\end{equation}
with i characterizing the proximity of the "sensor" with "sensor 1" in the tree structure, and where the charge transfer resistance \textit{R} and the electrolytic conductance \textit{G} are approximated as constants.\\
The different spectra are presented for three different sizes n for the network (Fig.~\ref{fig:fig8}d--f), considering \textit{R} = 12.5~M$\Omega$, \textit{G} = 10~$\upmu$S, $\upalpha$\textsubscript{0} = 0.75, \textit{Q}\textsubscript{0} = 0.5~$\upmu$F$\cdot$s\textsuperscript{$\upalpha$\textsubscript{0}-1}, $\upalpha$\textsubscript{1} = 0.9 and \textit{Q}\textsubscript{1} = 4~nF$\cdot$s\textsuperscript{$\upalpha$\textsubscript{1}-1}.\\
In the case of a single cell connecting two sensitive elements (Fig.~\ref{fig:fig8}d), "sensor 1" displays a single relaxation with \textit{Q}\textsubscript{1},$\upalpha$\textsubscript{1} because of the two serial CPDs.
In contrast, "sensor 2" displays two relaxations with \textit{Q}\textsubscript{0},$\upalpha$\textsubscript{0} and \textit{Q}\textsubscript{1},$\upalpha$\textsubscript{1} resulting from the serial combination of a wire without CPD and a wire with CPD.
The spectra are very distinct and both impedances show most of their differences between 10~mHz and 10~kHz both in magnitude and phase (Fig.~\ref{fig:fig8}d).\\
When n = 3 layers of electrochemical cells are used as a 2\textsuperscript{3}$\times$1, the number of impedance scenarios increases with the number of paths including the same number of wires, but either with or without CPD (these n + 1 scenarios are illustrated by different colors in Fig.~\ref{fig:fig8}a).
The resulting impedance spectra are represented in Fig.~\ref{fig:fig8}e.
Because the model takes into account only two different development stages for the CPDs, only two relaxations are evidenced in the phase diagram (Fig.~\ref{fig:fig8}e).
The spectra of "sensor 1" remain very distinct from all the others.
Moreover, we noticed a shift towards low frequency of the phase minima (maxima for --$\phi$) related to the numbers of grown CPDs on the path of connecting a given input to the common output.
If layer-to-layer interconnectivity is related to the input's spatial vicinity, the phase shift can be regarded as a metric for the collective activity of an input with its closest neighbors.
This is comparable to pixel crosstalk in sensing matrices having an inherent spread function,\cite{Jiang2018} so such a 2\textsuperscript{n}$\times$1 network responses would be intrinsically convolutive in space.\cite{Gu2018}\\
Raised to 23 layers, a 2\textsuperscript{23}$\times$1 network could connect over eight million sensing inputs, which is more than the number of olfactory receptor neurons in a human nose.\cite{Moran1982}
Despite the larger number of impedance states caused by the multiple layers, we noticed that the impedance of "sensor 1" is still highly differentiable among all the 2\textsuperscript{23} ones (Fig.~\ref{fig:fig8}f).
With a larger number n, the phase-shift for the main relaxation is broader on the spectra.
Spectra of "sensor 1" with the closest neighbor "sensor 2" are more similar, but keep a phase difference of more than 15 degrees at 1~Hz.
Despite the simulation showing that parallelizing CPDs in a several-layer stack should enable sensitive elements classification without silencing each of them in the crowd, many technical aspects should be highlighted.
First, the spatial convolution is an important property of these networks: To promote classification, sensitive materials should be grouped in an array based on the similarity of their responses.
Second, the impedance of an output to any sensitive element increases linearly with the number of layers (from approximately 20~M$\Omega$ at 10~mHz in Fig.~\ref{fig:fig8}d to about 0.5~G$\Omega$ at 10~mHz in Fig.~\ref{fig:fig8}f).

\newpage

Therefore, the voltage to apply must be increased accordingly to activate CPD growth: if 10~V\textsubscript{p} were required here for a 2$\times$1 elementary system, 230~V\textsubscript{p} would be necessary to enable the same structures in a 2\textsuperscript{23}:1 network.
To minimize the voltage to apply on such structures, alternative monomers,\cite{Koizumi2016,Baron2026a,Go2026} with a lower oxidation potential would be an asset.
In addition, reducing the gap between electrodes in a multilayer co-integrated system should also promote reducing the applied voltage, given the electric-field activation of this electrochemical phenomenon.\cite{Koizumi2016}

\section{Conclusions}
In this study, CPD morphogenesis was studied in systems that integrate sensitive elements and successfully showed that CPDs can store sensor activities in their morphology.
Connected to a serial resistor, a voltage supply must be increased, depending on how much the added impedance load promotes a voltage drop.
If this load is sensitive to its environment, a variation can trigger the growth of a CPD, specifically according to the nature of such a change on the effect of the load, such as the chemical composition of volatile compounds in air.
In cases where multiple loads coexist within the same growth medium, they may compete while evolving towards a common output. 
This behavior is driven by the different environmental influences acting on each respective impedance load.
Numerical simulations extended to a network of such systems indicate that, if this hardware can be fabricated, CPD morphogenesis could promote the calibration of an eNose to recognize user-specific environmental classes, even at a level where the dimension of a sensitive array overpasses the total number of olfactory receptor neurons humans have.
Such a passive element would considerably lower the energy cost necessary to manufacture sensing arrays but also the energy required to train an array to classify environmental information without remote resources.
A subsequent study on co-integrating such CPDs in a parallelized-node hardware will be reported soon, involving different materials and device geometries to optimize the voltage window triggering such CPD growths.

\section*{Acknowledgments}
This work is funded by ANR-JCJC "Sensation" project (grant number: \href{https://anr.fr/Projet-ANR-22-CE24-0001}{ANR-22-CE24-0001}).

\section*{Competing Interests}
The authors declare no competing interests.

\bibliographystyle{natsty-doilk-on-jour}  
\bibliography{ref}  

\newpage

\end{document}